\documentclass[aps,superscriptaddress,nofootinbib]{revtex4}



\usepackage{natbib}

\usepackage{graphicx}
\usepackage{dcolumn}
\begin{document}


\title{Trading activity and price impact in parallel markets: \\SETS vs. off-book market at the London Stock Exchange}



\author{Angelo Carollo}
\affiliation{Dipartimento di Fisica, Universit\`a di Palermo,
Viale delle Scienze, Ed. 18, I-90128, Palermo, Italy}

\author{Gabriella Vaglica}
\affiliation{Dipartimento di Fisica, Universit\`a di Palermo,
Viale delle Scienze, Ed. 18, I-90128, Palermo, Italy}

\author{Fabrizio Lillo}
\affiliation{Dipartimento di Fisica, Universit\`a di Palermo,
Viale delle Scienze, Ed. 18, I-90128, Palermo, Italy}
\affiliation{Santa Fe Institute, 1399 Hyde Park Road, Santa Fe, NM 87501, U.S.A.}
\affiliation{Scuola Normale Superiore di Pisa, Piazza dei Cavalieri 7, 56126 Pisa, Italy}

\author{Rosario N. Mantegna}
\affiliation{Dipartimento di Fisica, Universit\`a di Palermo,
Viale delle Scienze, Ed. 18, I-90128, Palermo, Italy}
\date{\today}

\begin{abstract}
We empirically study the trading activity  in the electronic on-book segment and in the dealership off-book segment of the London Stock Exchange, investigating separately the trading of active market members and of other market participants which are non-members. We find that (i) the volume distribution of off-book transactions has a significantly fatter tail than the one of on-book transactions, (ii) groups of members and non-members can be classified in categories according to their trading profile (iii) there is a strong anticorrelation between the daily inventory variation of a market member due to the on-book market transactions and inventory variation due to the off-book market transactions with non-members, and (iv)  the autocorrelation of the sign of the orders of non-members in the off-book market is slowly decaying. We also analyze the on-book price impact function over time, both for positive and negative lags, of the electronic trades and of the off-book trades.   The unconditional impact curves are very different for the electronic trades and the off-book trades. Moreover there is a small dependence of impact on the volume for the on-book electronic trades, while the shape and magnitude of impact function of off-book transactions strongly depend on volume.
\end{abstract}


\maketitle


\section{Introduction}

Today an investor willing to trade a given security faces some choices when he or she decides where to perform the trade. In fact, she has the possibility of trading in many different trading venues that differ in structure, rules, fee structure, etc. A major distinction between trading venues is the one between automated  and intermediated  market segments. In the former type there is typically a centralized electronic order book where orders are submitted and executed automatically, while in the latter type intermediaries perform their trades on a bilateral basis. Usually, this segment of the market (the off-book or upstairs segment) is preferentially (but not exclusively) used by large institutional traders.  The interaction between the trading in different market segments is still relatively poorly explored as testified by the documents asking for opinion published in earlier 2010 by the Security and Exchange Commission (SEC) \cite{SEC10} and by Committee of European Securities Regulators (CESR) \cite{Cesr10}. In particular it is important to understand how different market segments interact one with each other through the trading of investors and their chosen strategy.

The literature on multimarket trading includes studies of the choice between (i) regional exchanges and primary exchanges \cite{Easley1996,Bessembinder1997}, (ii) purely electronic platforms (ECNs) and traditional market makers \cite{Barclay2003}, and (iii) high liquidity and low liquidity market segments of the U.S. Treasury securities  \cite{Barclay2006}. Theoretical descriptions of these segments within a market can be found in \cite{Seppi1990,Grossman1992}. Empirical investigations of the economic benefits of the market fragmentation in terms of spread, price impact  and adverse selection have been performed in several markets in different years. Examples are the studies describing the nature of trades in the downstairs and upstairs segments of the NYSE \cite{keim,Madhavan1997,Madhavan1998}, the investigation of adverse information and price effects in the segments of the Toronto Stock Exchange \cite{Smith2001}, the investigation of the permanent and temporary price impact of the Helsinki Stock Exchange \cite{Booth2002}, the study of the information content of upstairs trades in the Paris Bourse \cite{Bessembinder2004}, and the study of order flow interactions between the on-book and off-book segments of the London Stock Exchange (LSE)  \cite{jain,Dufour2005,friederich}.

When similar studies are performed one key aspect concerns the access to data. In fact, in the present market structure it is very difficult to access data containing a full coverage of all the market venues and all market segments. Moreover even if transaction data are available for all the market segments, these data do not usually allow to follow individually the trading activity of investors, brokers, market members, etc., because trade and quote data are often anonymized and the tendency of the spreading of anonymity is growing \cite{Comerton2009}.

In this paper we want to empirically investigate the trading activity of market participants in a twofold segment market structure, namely the London Stock Exchange (LSE) by following the all trading decisions of registered market participants performed during 2004. In that year LSE had a dual structure made of an electronic order driven market ({\it on-book market}) and a bilateral dealership market ({\it off-book market}) where market members can trade with other members or with other participants. The database we investigate is the Rebuild Order Book of the LSE. In our version the database contains the coded information of the counterparts involved in each on-book or off-book transaction. The codes identifying participants trading at LSE are associated with stock exchange market members for those trades occurring in the on-book section of the market. In the case of off-book transactions some codes are associated with market members while others refer to market participants which are not members of the LSE. We will shortly address these participants as ``non-members". In our study we therefore investigate the inventory variation of a certain number of stocks of most active market members to detect stylized facts in their trading activity and market impact in the on-book and off-book market sections. We also investigate statistical regularities observed for market participants which are not members and are therefore active only in the off-book market segment. This second analysis will be performed at an aggregate level because we assume that a large number of investors are associated with the same ``non-member" code.

The use of market member data to infer trading strategies is relatively new and has been recently explored, for example, in Ref. \cite{Lillo2008,Vaglica2008,Zovko2007,Moro2009,Vaglica2010}. It has been shown that these codes provide useful information to characterize the strategic behavior of investors, mainly because market members either act on their own behalf, or they trade often on behalf of a large trader, or because a market member often specialize on a class of final investors. In this latter case we can consider the member strategy as an effective strategy resulting from the aggregation of a relatively homogeneous set of investors. The main purpose of this paper is to identify new statistical regularities of the trading activity of market members and market participants acting in different market segments and to characterize the price impact of transactions happening either on-book or off-book.

 The paper is organized as follow. In Section II we describe the market structure and the data that we are using. In Section III we show some results about the volume distribution of individual transactions in different market segments. In Section IV we investigate the trading activity of market members and in Section V we consider some statistical properties of the trading activity of non-members. In Section VI we show the difference of price impact of on-book and off-book transactions and Section VII concludes.

\section{Market structure and data}

Our empirical analysis is based on data of the London Stock Exchange (LSE) recorded in 2004.
LSE gives market participants the possibility to trade domestic UK stocks in different venues (SETS, SETSmm, SEAQ, or SEATS),
The SETS (Stock Exchange Trading System), introduced in the 1997, is the only completely order driven market, in which no market makers are required. Bids and
offers are determined by orders that exist in the electronic orderbook and are automatically matched. High capitalized stocks including FTSE
100 are traded in a segment of  SETS termed SET1. Market participants can be divided in two classes, market members and non-members. Membership is usually available to
investment firms and credit institutions authorized in the European Economic Area. Market participants who are not members of the LSE are of two kinds, retail and
institutional investors. Retail investors are small firms and individual investors, while institutional investors include mutual funds, hedge funds, pension funds, etc.
Only the members of the LSE have direct access to SETS. Other market participants (e.g. non-members) can exchange securities only through the market members in essentially two different ways. They can access the SETS through the agency services of members or access directly the dealership market exchanging securities with members. Therefore an individual or an institution who wishes to trade on the LSE must contact his/her broker/dealer and instruct him/her to submit an order to the SETS order book or can negotiate a trade with the dealer as a counterpart.

Most of the members offer agency services and at the same time do business as proprietary trades in the electronic (SETS) market and in the dealership off-book market. Often the agency service and proprietary investment activities are two completely separated areas inside the same firm. For this reason it is not uncommon to see transactions in which the buyer and the seller are the same market member.

Private investors generally have the possibility to choose the most appropriate brokerage service depending on their characteristics, like the amount of money at their disposal, the knowledge of the stock market, the time they want to allocate for the investment and so on. There are some differences regarding the access to the securities between retailer and institutional investors. Retail investors access the market through their broker, often using online brokering services. At LSE there is a group of member firms, the Retail Service Providers (RPS), specialized in the small orders, who supplies the retail order execution systems and guarantees prices at or inside the spread. Usually the retail investors, through the brokers, contacts the RPS and obtain prices which are often better than those of SETS. For this reason most trading done by UK retail investors is conducted off-book with RPS.
Institutional investors cannot access SETS directly as well but can place orders directly with a member firm via telephone or via computer. In this case, the member firm can fill the order in several ways: from its own inventory, through a customer matching, routing the order to another member firm, executing the order via SETS for the client acting as agent, giving its customer a Direct Market Access (DMA) offering to the client a direct pipeline to SETS (for sophisticated private investors only), and finally doing a Contract for Difference (CDF)\footnote{In this last case the member  trades a security on its own behalf and enters into a CDF with the clients, meaning that no shares are exchanged between them but the client assumes the risk associated with handling the shares.}. In the first three cases the transaction is off-book while in the last three cases the transaction is on-book.

The data-set we investigate is the Rebuild Order Book of LSE. Specifically, we investigate nine highly liquid stocks, Vodafone GPR (VOD), GLAXOSMITHKLINE (GSK), BP (BP), LLOYDS TSB GRP (LLOY), ROYAL BANK SCOT (RBS), HSBC HLDGS.UK (HSBA), SHELL (SHEL), ASTRAZENECA (AZN), BT GROUP (BT-A), traded during the whole year 2004. The analysis is performed separately for each stock. The Rebuild Order Book contains information both on on-book and off-book transactions. Specifically, the on-book transaction and order records include prices, volumes, and submission, transaction or cancellation time. For each order  our version of the database contains the anonymized identity of the market participant who placed the order. In the case of off-book transactions, the database contains price, volume, and time of the transactions and the anonymized identity of the two counterparts. For off-book trades the counterparts can be either market members or non-members. With our data we are therefore able to completely follow -- (i) the trading activity of market members, acting as principal and/or agency capacity, within each securities in the on-book market and as principal or riskless principal in the off-book market and (ii) the trading activity of non-members in the off-book market. It is worth to note that Ref. \cite{Dufour2005} finds that the vast majority of  trades executed on SETS are between members acting in principal capacity.
In analogy to what has been done in Ref. \cite{Dufour2005}, we consider here only ordinary trades. Ordinary trades are identified as trade type \verb"AT" and \verb"O" for the on-book and off-book trades, respectively. On average AT trades constitutes more than 95\% of the transactions and more than 90\% of volume of the on-book market. Similarly O trades constitutes more than 95\% of the transactions and more than 85\% of the volume of the off-book market.

We do not have direct information about which codes correspond to market members and which to non-members.
We know that each market member has a different code, while we do not know whether distinct investors acting as non-members share the same code.
We use an heuristic approach to classify a code as belonging to a market member or to a non-member. Market members are the only ones having the right to trade in the on-book section of the market. We therefore describe as non-members the market participants not acting in the on-book section of the market during 2004. Specifically, we classify as non-members all codes for which we do not find any electronic trade in the on-book segment of the market for all of the 92 stocks of SET1 during the entire 2004. Accordingly, we classify as market members all the remaining codes which are not classified as non-members, i.e. all codes for which we find at least one electronic on-book transaction in at least one of the 92 stocks of SET1 during 2004.
Even if this classification cannot be guaranteed as error free, we are pretty confident that it correctly classifies the vast majority of the LSE market participants. In total we find $229$ distinct market member codes and $163$ distinct non-member codes. These numbers suggest that many different investors are described by the same non-member code. This is certainly true for retail investors and it is probably true also for institutional investors acting as market participants in the off-book market segment.

We know that the identity of the market member is coherently coded in our database and therefore we can (i) analyze their trading activity individually, (ii) disaggregate the contribution coming from the on-book and from the off-book transactions to their inventory variation, and (iii) compute the correlation between the inventory variation of each pair of market members. The process of coding of non-members acting in the off-book is less well defined and therefore we decide to perform  only a study of their aggregated trading behavior for these market participants.

The classification of market participants in effective market members and non-members allows us to introduce a classification of transactions done on the basis of the market venues and of the nature of the two counterparts.
Specifically, in the present paper we consider four mutually exclusive sets of transactions: the on-book transactions occurring between market members (OnMM), the off-book transactions occurring between market members (OffMM), the off-book transactions occurring between a market member and a non-member (OffMN), and the off-book transactions occurring between two non-members (OffNN).

\section{Size distribution of  trades in different market segments}

\begin{figure}[tH]
\includegraphics[scale=0.6]{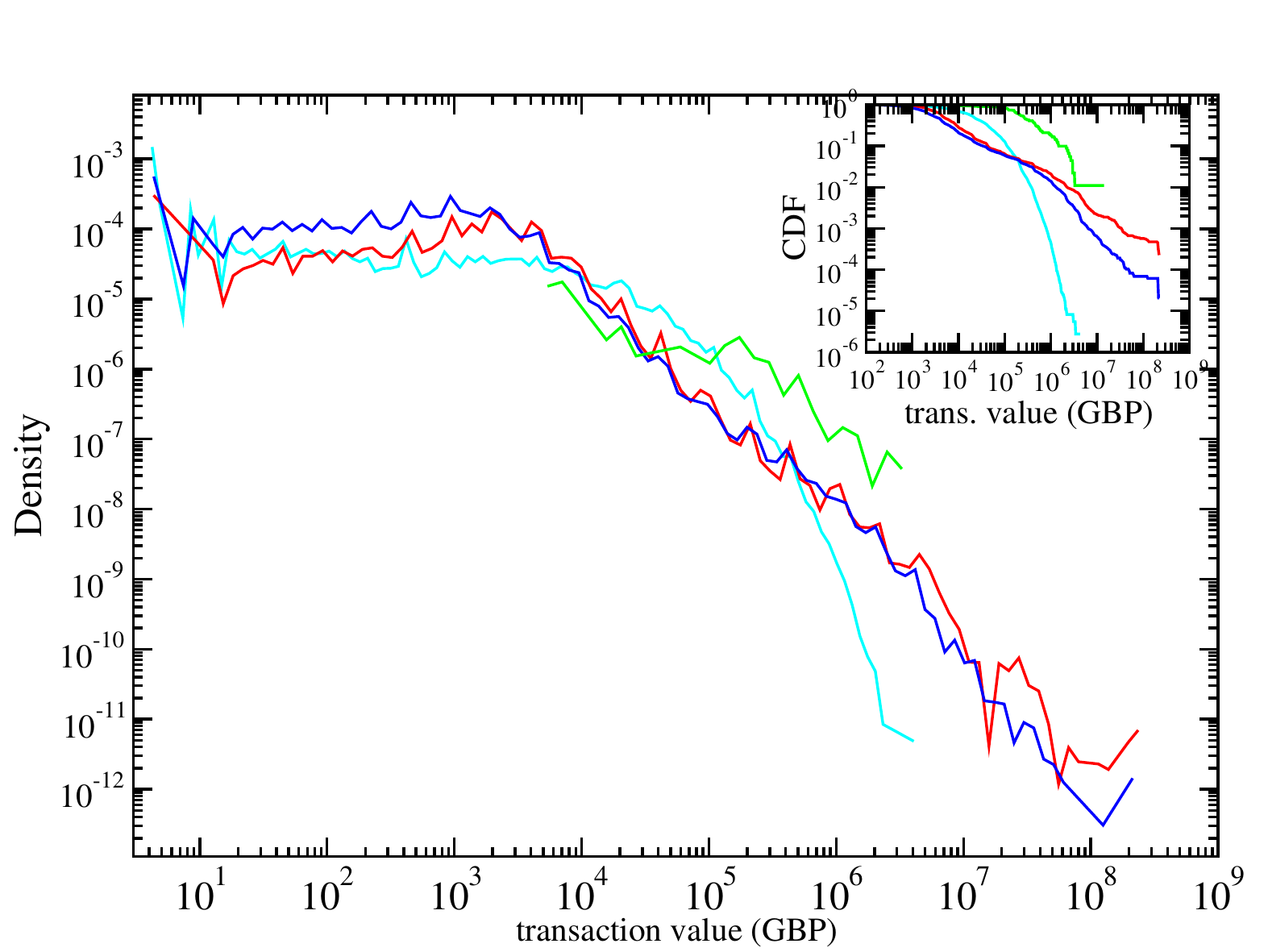}
\caption{\small{Probability density function of the transaction value given in GBP for four different types of market transactions. The cyan curve is the PDF of value of the on-book transactions (set OnMM), the red curve is for the off-book transactions between two members (set OffMM), the blue curve for the off-book transactions between a member and a non-member (set OffNM) and the green curve for the off-book transactions between two non-members (set OffNN). The traded stock is LLOY. The inset shows the cumulative probability density function of the transaction value given in GBP for the 4 different sets of transactions. }}
\label{fig:1}
\end{figure}

We consider first the distributional properties of single transaction value for the four sets of transactions defined in the previous section. The value of a transaction is defined as the product of the transaction price times the exchanged volume.
We choose to use this variable instead of exchanged volume to compare the trading activity of different stocks that might be characterized by different levels of price per share.
Table \ref{summary} presents a summary statistics of the number of transactions and exchanged value for the four sets of trades. In the case of exchanged value we also report its average, standard deviation, median and maximum value.
The Table shows that while the percentage of on-book transactions varies between a maximum value of $83.5\%$ for (AZN) and a minimum of $56.9\%$ for BT-A, the percentage of exchanged value of these trades varies only between  a maximum value of $57\%$ (HSBA) and a minimum value of $40\%$ (BT-A). Roughly $64\%$ of off-book transactions are between a member and a non-member and only a tiny fraction is between two non-members although these transactions are on average very large in exchanged value.

\begin{table}
\begin{center}
\caption{{\small Summary statistics of the transactions for the nine stocks investigated during 2004. The table is relative to the transactions labeled by the LSE as transactions of type AT and O both in the on-book electronic market and in the off-book market. We investigate the number of transactions (NoT), the total exchanged value $V_{tot}$ in all considered transactions, the average exchanged value for transaction Av($v$), its standard deviation SD($v$), its median Md($v$) and its maximum value Max($v$). These quantities are computed for the four sets of transactions indicated as (i) OnMM, (ii) OffMM, (iii) OffMN and (iv) OffNN, respectively.  The total exchanged value $V_{tot}$ is given in billion GBPs whereas the average exchanged value per transaction Av($v$), its SD($v$), Md($v$) and Max($v$) are given in million GBPs.}}
\vskip 0.5cm
\begin{small}
\begin{tabular}{|l||r|r|r|r|r|r|r|r|r|} \hline
&VOD& GSK& BP&LLOY& RBS& HSBA& SHEL& AZN& BT-A\\
\hline
NoT i&426,159&413,213&440,526&365,175&398,137&380,422&386,500&416,670&220,500\\
NoT ii&75,775&66,875&61,035&84,507&65,668&41,589&81,914&26,730&65,376\\
NoT iii&146,084&92,777&105,247&146,193&89,226&99,593&107,102&55,366&101,693\\
NoT iv&235&144&172&91&115&151&123&119&106\\
\hline
$V_{tot}$ i&48.871&31.967&41.303&17.403&26.972&39.486&22.965&24.377&10.509\\
$V_{tot}$ ii&16.816&18.952&23.590&20.256&8.757&15.161&15.307&8.186&9.210\\
$V_{tot}$ iii&20.916&15.049&18.308&11.692&12.892&14.95&10.713&11.452&6.561\\
$V_{tot}$ iv&0.8396&0.9082&0.8844&0.0651&0.08323&0.1687&0.1098&0.0871&0.068\\
\hline
Av($v$) i&0.115&0.0774&0.0938&0.0477&0.0677&0.104&0.0594&0.0585&0.0477\\
Av($v$) ii&0.222&0.283&0.386&0.240&0.133&0.364&0.187&0.306&0.141\\
Av($v$) iii&0.143&0.162&0.174&0.08&0.144&0.15&0.10&0.207&0.0645\\
Av($v$) iv&3.573&6.307&5.142&0.716&0.724&1.117&0.893&0.732&0.645\\
\hline
SD($v$) i&0.317&0.142&0.158&0.079&0.116&0.188&0.095&0.096&0.093\\
SD($v$) ii&2.389&5.572&6.635&5.204&1.184&1.933&3.746&2.143&3.308\\
SD($v$) iii&0.917&1.213&2.066&1.89&1.30&1.232&0.822&1.058&1.183\\
SD($v$) iv&21.193&37.273&33.905&1.618&1.080&1.439&1.524&0.920&1.247\\
\hline
Md($v$) i&0.0227&0.0341&0.0414&0.0216&0.0336&0.0374&0.0291&0.0289&0.0182\\
Md($v$) ii&0.0050&0.0058&0.0062&0.0049&0.0061&0.0070&0.0059&0.0084&0.0024\\
Md($v$) iii&0.0030&0.0050&0.0041&0.0033&0.0039&0.0046&0.0037&0.0047&0.0016\\
Md($v$) iv&0.593&0.305&0.446&0.265&0.307&0.704&0.305&0.425&0.281\\
\hline
Max($v$) i&12.763&6.009&5.34&4.357&7.438&4.553&4.083&4.689&3.61\\
Max($v$) ii&280.5&357.75&389.2&257.767&116.25&76.146&233.4&153.24&253.012\\
Max($v$) iii&89.601&255.36&367.56&233.2&173.34&143.36&113.4&42.064&163.8\\
Max($v$) iv&212.963&269.75&295.9&14.24&6.47&8.9&12.714&4.68&9.781\\
\hline
\end{tabular}
\label{summary}
\end{small}
\end{center}
\end{table}

Figure \ref{fig:1} shows the probability density function (PDF) of  transaction value for the four different types of transactions of the stock LLOY. We choose to use LLOY throughout the paper to present our results. However, unless when explicitly stated differently, the behavior observed for LLOY is representative of all nine investigated stocks.
The PDF of the on-book transaction value (cyan curve of the figure) is quite flat in the range $10^1\div 10^4$ GBP. For values larger than $10^4$ GBP the tail of the PDF decays quickly to zero. On the contrary both off-book transactions between two market members (set OffMM, red curve) and between a market member and a non-member (set OffNM, blue curve) have a transaction value PDF characterized by similar fat tails. The two PDFs differ for transaction value smaller that $10^3$ GBP. Below this value, OffMN transaction value PDF is larger than OffMM transaction value PDF. This is probably due to the trading activity of retail investors that are accessing the market as non-members. The PDF of the transaction value of the set OffNN (green curve) is also fat tailed, but the extension of the fat tail is quite limited due to the small number of events of this set. In fact, the sample is composed by less than one hundred elements.


The pronounced tail of the transaction value PDFs suggests an asymptotic power law behavior, especially in the cases of off-book transactions between two market members and between a market member and a non-member. An asymptotic power law behavior has been observed in Ref. \cite{gopi2000} when investigating US equity data. In their study, authors find an exponent $1.5$ for the tail exponent of the cumulative distribution of transaction volume. It is worth pointing out that in their analysis of the transaction volume PDF authors are investigating together both the on-book and the off-book transactions (for an economic interpretation of this result see \cite{gabaix2006}). The present investigation shows that the distributional properties of the value of transactions occurring at the on-book and off-book segments of the LSE market are quite different. Specifically, a power law behavior with an exponent close to the value observed when investigating US equity data is observed only for the off-book transactions. As originally suggested in Ref. \cite{lmf2005}, we hypothesize that, in the US equity data, the power law tail of transaction volume distribution is probably due primarily to the off-book transactions.

We quantitatively characterize the power law behavior of the tail of the transaction value PDF by computing the 1\% Hill estimator of the tail exponent of the cumulative probability density function. The results are shown in Table \ref{Hill}. On average (performed across the different stocks) the cumulative distribution of off-book transactions value between two members has a tail exponent of $1.44$ and the cumulative distribution of off-book transactions between a member and a non-member has a tail exponent of $1.59$. On-book transaction value cumulative distribution is characterized by  quite different 1\% Hill estimator. In fact its average value is  $2.85$. Moreover,  as it can be seen from the profile of the cumulative distribution of value (cyan curve) shown in the inset of Fig. \ref{fig:1}, the 1\% top quantile is indeed not describing a power law regime for on-book transaction. This visual observation is quantitatively supported by the fact that the $0.1\%$ Hill estimator (see Table \ref{Hill}), which locally better fits the tail behavior for largest on-book transaction values gives an average exponent of  $3.81$ which is quite different from the 1\% average value of $2.85$.

In conclusion, in agreement with previous observation reported in \cite{lmf2005}, we observe that on-book transaction values are described by a thin tailed distribution, i.e. either an asymptotically power law cumulative distribution with an exponent larger or close to 3 or a non power law distribution. On the contrary both sets of off-book transactions (OffMM and OffMN) are described by power law tail of the cumulative distribution of transaction value with a tail exponent close to $1.5$ extending on several orders of magnitude.

\begin{table}
\begin{center}
\caption{{\small Hill estimator of the tail exponent of the transaction value cumulative distributions for the nine investigated stocks. All exponents are given together with the $95\%$ of confidence interval. The threshold used in the estimation is in the table header. 
}}
\vskip 0.5cm
\begin{small}
\begin{tabular}{|l|c|c|c|c|} \hline
& OffMM 1\%& OffMN 1\% & OnMM 1\% & OnMM 0.1\%\\
\hline
VOD&1.50 $\pm$ 0.09&1.69 $\pm$ 0.07&2.24 $\pm$ 0.06&4.21 $\pm$ 0.34\\
GSK&1.20 $\pm$ 0.08&1.87 $\pm$ 0.10&2.43 $\pm$ 0.06&3.31 $\pm$ 0.28\\
BP&1.18 $\pm$ 0.08&1.60 $\pm$ 0.08&3.02 $\pm$ 0.08&4.01 $\pm$ 0.32\\
LLOY&0.87 $\pm$ 0.05&1.27 $\pm$ 0.06&3.15 $\pm$ 0.09&4.10 $\pm$ 0.36\\
RBS&1.75 $\pm$ 0.12&1.84 $\pm$ 0.10&2.59 $\pm$ 0.07&2.89 $\pm$ 0.25\\
HSBA&2.10 $\pm$ 0.17&1.57 $\pm$ 0.08&3.19 $\pm$ 0.09&4.56 $\pm$ 0.40\\
SHEL&1.16 $\pm$ 0.07&1.49 $\pm$ 0.08&2.86 $\pm$ 0.08&4.39 $\pm$ 0.38\\
AZN&1.85 $\pm$ 0.19&1.67 $\pm$ 0.12&2.86 $\pm$ 0.08&3.21 $\pm$ 0.27\\
BT-A&1.09 $\pm$ 0.07&1.30 $\pm$ 0.07&2.92 $\pm$ 0.11&3.59 $\pm$ 0.41\\
\hline
\end{tabular}
\label{Hill}
\end{small}
\end{center}
\end{table}

\section{Trading activity of members in the on-book and in the off-book market}

In this section we investigate the statistical properties of the trading activity of the market members actively trading in the on- and in the off-book segments of the LSE market. Specifically, we are interested in describing the inventory variation profile of a market member in the two market segments and in doing a statistical modeling of the synchronous inventory variation dynamics of all the market members which are active in a given market segment.

The number of distinct market members trading a specific stock is on average 192.6 and ranges from the minimum value of 181 to the maximum value of 202. The amount of trading  across market members is highly heterogeneous at LSE. Few members are responsible of a large fraction of the market transactions and of the exchanged volume. We quantify this heterogeneity by using the Gini coefficient \cite{Gini}. Given a distribution of weights among a certain number of elements (in our case the market members), a Gini coefficient equal to 0 indicates a perfect equality in the distribution of weights among the elements while a value of the coefficient equal to 1 indicates a total inequality of the distribution,  i.e.  only one element has a non vanishing weight. For each stock we measure the Gini coefficient of the fraction of market transactions and of the fraction of exchanged value by the market members trading the investigated stocks in the two market segments. The Gini coefficient of on-book fraction of market transactions is on average $g_N^{on}= 0.881 \pm 0.005$ and the one for fraction of exchanged value is $g_V^{on}=0.878 \pm 0.006$. For off-book trading the corresponding values are $g_N^{off}=0.921 \pm 0.009$ and $g_V^{off}=0.897 \pm 0.01$ for the fraction of transactions and for the fraction of exchanged value, respectively. We observe that these values are very high and close to one, indicating a strong heterogeneity in trading activity.  We notice than within the on-book market the Gini coefficient for the fraction of transactions and the exchanged value are not statistically different, since a t-test cannot reject the hypothesis that the mean value of $g_N^{on}$ and $g_V^{on}$ are the same. On the other hand, a t-test shows that $g_N^{off}$ and $g_V^{off}$ are statistically different, indicating that market members in the off-book are more heterogeneous with respect to the fraction of transaction than to the exchanged value. These two observations together can be summarized by saying that the heterogeneity of market members' trading activity is very high and it is slightly more pronounced in the off-book than in the on-book market.


\begin{table}
\begin{center}
\caption{{\small Number of members in the three investigated sets (A, B, and C) for the nine stocks. Set A includes members active mainly in the on-book market, while set C include members mainly active off-book. Set B include those members very active in both segments.}}
\vskip 0.5cm
\begin{small}
\begin{tabular}{|l|c|c|c|} \hline
&Set A&Set B&Set C\\
\hline
VOD& 45 & 40 & 38 \\
GSK&43 &38 &35 \\
BP&43 & 38& 33 \\
LLOY&52 &32 & 39\\
RBS&40 & 38&34 \\
HSBA&41 &37 &32 \\
SHEL&48 &33 &41 \\
AZN&36 &42 &27 \\
BT-A&46 &32 &24 \\
\hline
\end{tabular}
\label{members}
\end{small}
\end{center}
\end{table}

The large heterogeneity of market member trading activity implies that a significant fraction of members can be quite inactive for long periods  of time. The present study aims to track the inventory variation of very active market members. For this reason, for each stock and for each market segment we select the most active market members by requiring that their cumulative trading activity is responsible of $99\%$ of total transactions occurred for the considered stock in the considered market segment. The intersection of the two sets of the most active members on- and off-book is termed set B. The difference between the set of most active members on-book and set B is termed set A, while the difference between the set of most active members off-book and set B is termed set C.
Table \ref{members} shows the number of market members of set A, B, and C for the 9 investigated stocks. It is worth noticing that the size of the three sets is roughly equal, indicating that market members can be divided in three subsets of similar size: those specialized in trading in the on-book market (set A), those specialized in trading in the off-book market (set C), and those active in both segments (set B).

\begin{figure}[Ht]
\includegraphics[scale=0.5]{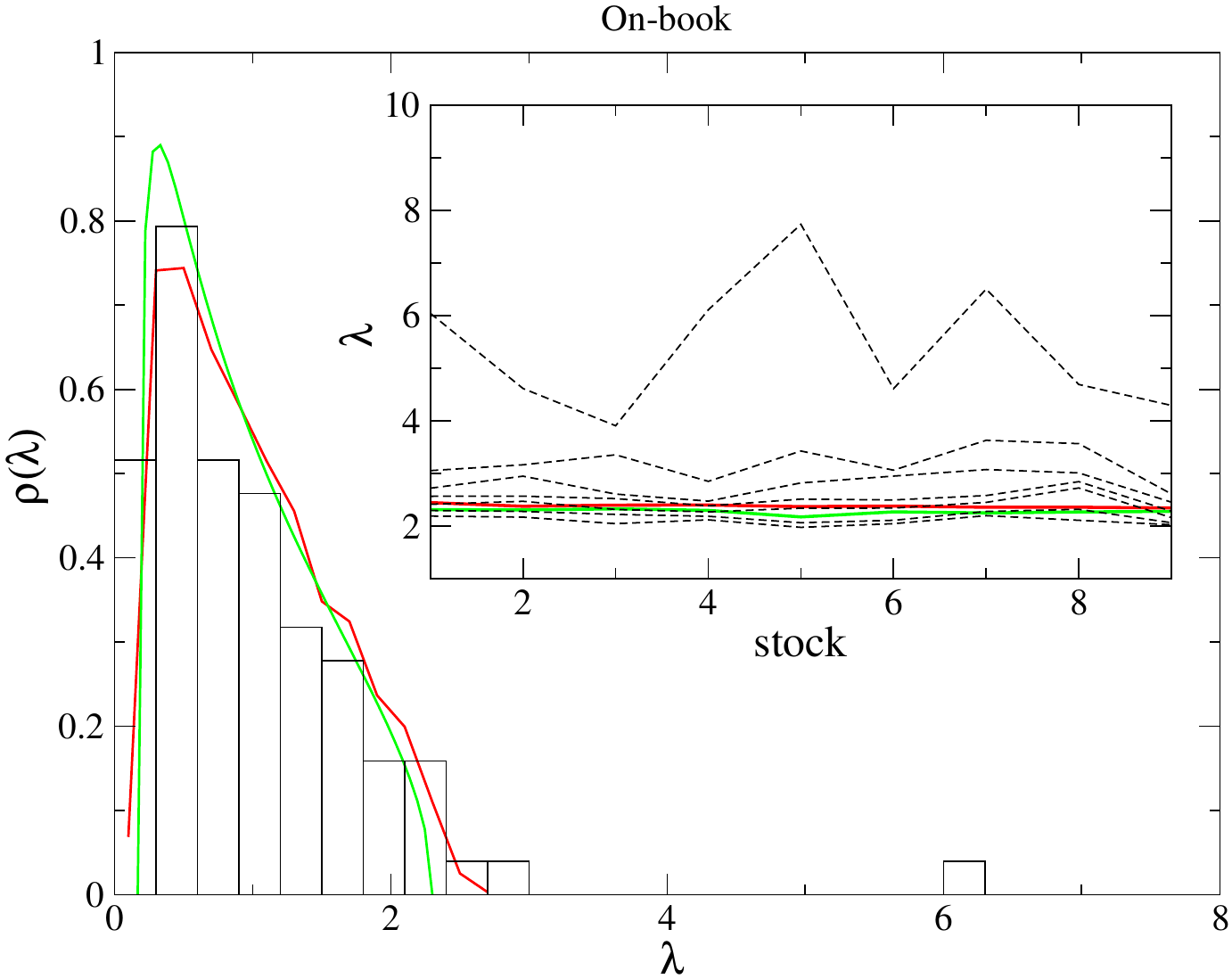}
\includegraphics[scale=0.5]{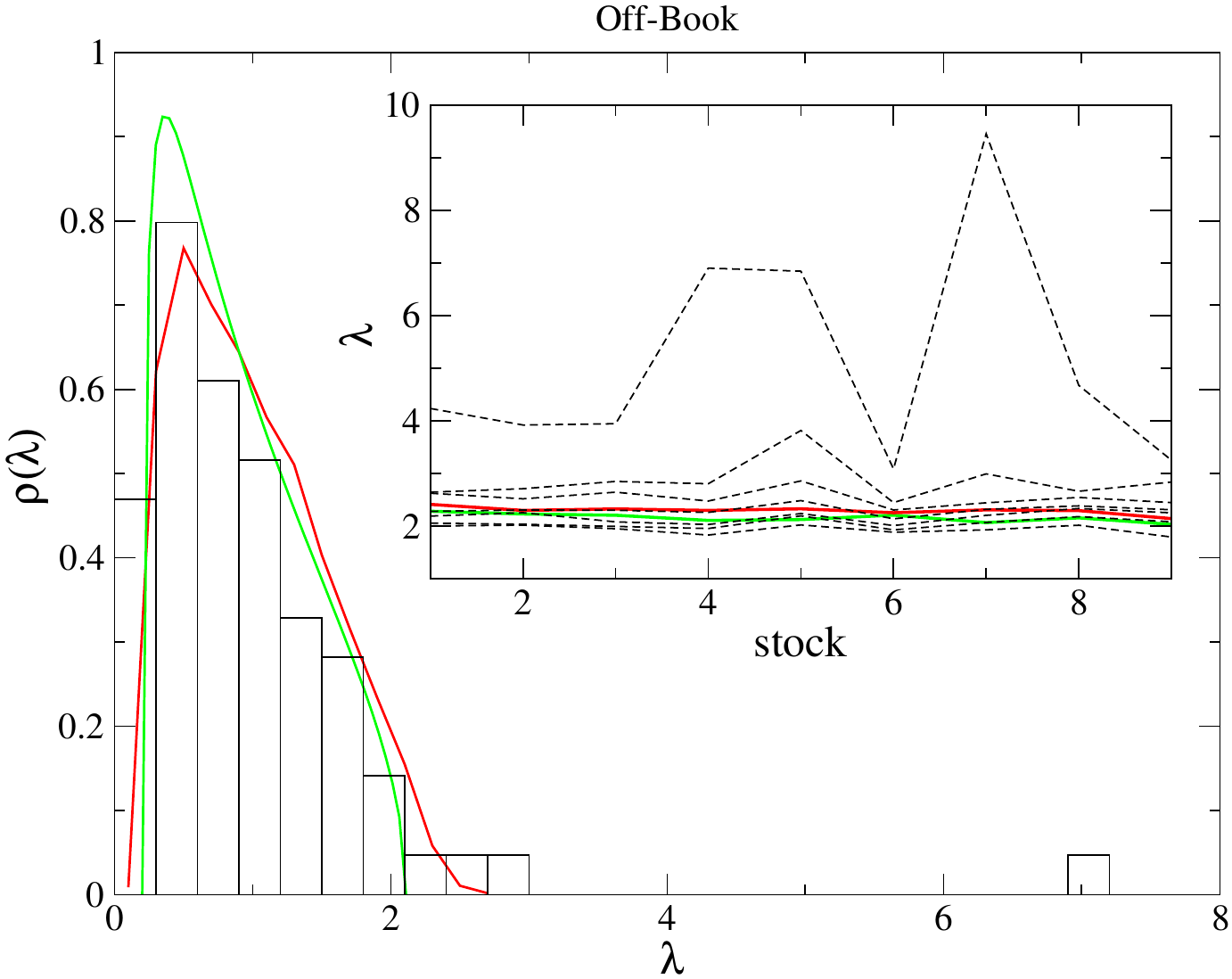}
\caption{Histogram of the eigenvalues  of the correlation matrix of the inventory variations of the market participants trading actively on the stock LLOY in the on-book (set A and set B, left) and off-book (set B and set C, right) market. The green line is the spectral density predicted by random matrix theory where the contribution of the first eigenvalue is removed. The solid red line is the averaged spectral density obtained by shuffling independently the time series of each participants.
Each inset shows the 6 largest eigenvalues (black dashed lines) of the correlation matrices of each of the 9 stocks considered (VOD, GSK, BP, LLOY, RBS, HSBA, SHEL, AZN and BT-A). The green solid lines are the corresponding largest eigenvalues of the spectral density predicted by random matrix theory. The red solid lines are the averaged largest eigenvalues obtained by the shuffling procedure.}
\label{PCAfig}
\end{figure}

In order to quantify the trading activity of a market member on a given stock, we consider its daily inventory variation in GBP. This is defined as the value exchanged by the market member as a buyer minus the value exchanged as a seller on the selected stock in a day. For market members of set A we investigate separately the on-book inventory variation time series, for members of set C we investigate the off-book inventory variation time series, while for members of set B we investigate inventory variation time series  associated with on- and off-book transactions.

The similarity of the inventory variation time profile of a pair of market members acting in a given market segment can be measured by the linear cross correlation between their inventory variation in that segment. Therefore for each market segment, the similarity of trading activity across market members in that segment is captured by the cross correlation matrix of inventory variations.

In Ref. \cite{Lillo2008} three of us studied the cross correlation matrix of inventory variation of market members trading in the on-book market segment of the Spanish Stock Exchange (BME)
and found that a large number of market members can be classified in two groups. Market members belonging to the same group are characterized by a positive correlation of inventory variation among them, whereas a negative correlation is observed between the inventory variation of market members belonging to distinct groups. The spectral analysis of the correlation matrix shows the presence of a relatively large eigenvalue associated with a factor driving a significant fraction of  the overall dynamics of inventory variation. In Ref. \cite{Lillo2008} authors have shown that this factor is highly and significantly correlated with price return of the traded stock.

We perform here a similar analysis on the LSE data in order to verify whether these findings are observed also in this market and in order to investigate them in different market segments.
For both the on-book and the off-book market segments the correlation matrix of inventory variation has both positive and negative statistically correlation coefficients which are reliable at a $2\sigma$ threshold level.
As in \cite{Lillo2008}, in order to investigate whether the on-book and off-book inventory variations carry information on the market dynamics, we perform a principal component analysis of each cross-correlation matrix computed in each market segment. We estimate the number of principal components which are not affected by the unavoidable statistical uncertainty associated with the finite number of records used to estimate the correlation matrix by using results from random matrix theory \cite{Laloux1999,Plerou1999,Plerou2002}. The underlying idea is that comparing the eigenvalue distribution of a correlation matrix with that obtained from a set of uncorrelated variables provides a way to identify relevant principal components, as those that cannot be explained by statistical uncertainty associated with correlation coefficients estimation. Specifically,  eigenvalues exceeding the value of the maximum eigenvalue expected according to random matrix theory are assumed to carry information on market dynamics.
Random matrix theory assumes as a null hypothesis that inventory variations are independent Gaussian time series with finite variance. However real inventory variations might have different statistical properties and are also subjected to a constraint by the fact that the sum of inventory variations of a given day across all the members must be zero (at least in the on-book segment). For this reason, in analogy with Ref. \cite{Lillo2008} we supplement the random matrix theory theoretical results with the result of numerical simulations performed by shuffling the time series of inventory variations of every participant.

Figure \ref{PCAfig} shows the spectral density of the inventory variation cross correlation matrix for LLOY in the on-book (left) and the off-book (right) market segment. In both cases the first eigenvalue has a value much larger than the one expected in terms of random matrix theory or observed in the shuffling experiments.
Specifically, the first eigenvalue of the correlation matrix of inventory variation for the market members active in the on-book market market segment (i.e. union of the market members belonging to set A and set B) explains, on average, $6.7\%$ of total variance. This value should be compared with the correspondent $12.8\%$ observed for the BME market in \cite{Lillo2008}. An analogous comparison for the percent of variance explained by the second eigenvalue relates the $4.0\%$ of the total variance observed at LSE with  the $4.3\%$ observed at BME. For the market members active in the off-book market segment (i.e. union of market members of set B and set C),  the first (second) eigenvalue explains, on average, the $7.3\%$ ($4.1\%$) of total variance.  As in \cite{Lillo2008}, the first factor of the principal component analysis of the on-book inventory variation cross-correlation matrix is highly correlated with the daily price return of the traded stock, calculated as the difference between the logarithm of closing price and the logarithm of opening price. The correlation between price return and the first factor of the correlation matrix of on-book inventory variation is, on average,  $0.55$, and it is ranging from $0.37$ for AZN to  $0.67$ for GSK. For off-book trading the first factor is again correlated with the price return of the traded stock, and the correlation is, on average, $0.31$. The range of variation being from the minimum value of $0.15$ for BT-A to the maximum value of $0.45$ for LLOY\footnote{It is worth noting t that the inventory variation time series in the on-book and in the off-book market segments are computed in a slightly different way. The daily on-book inventory variation is obtained by considering transactions which are occurring during market hours whereas daily off-book inventory variation include transactions which might occurs after market hours although of course before the next opening of the market.}.


\begin{figure}[Ht]
\includegraphics[scale=0.8]{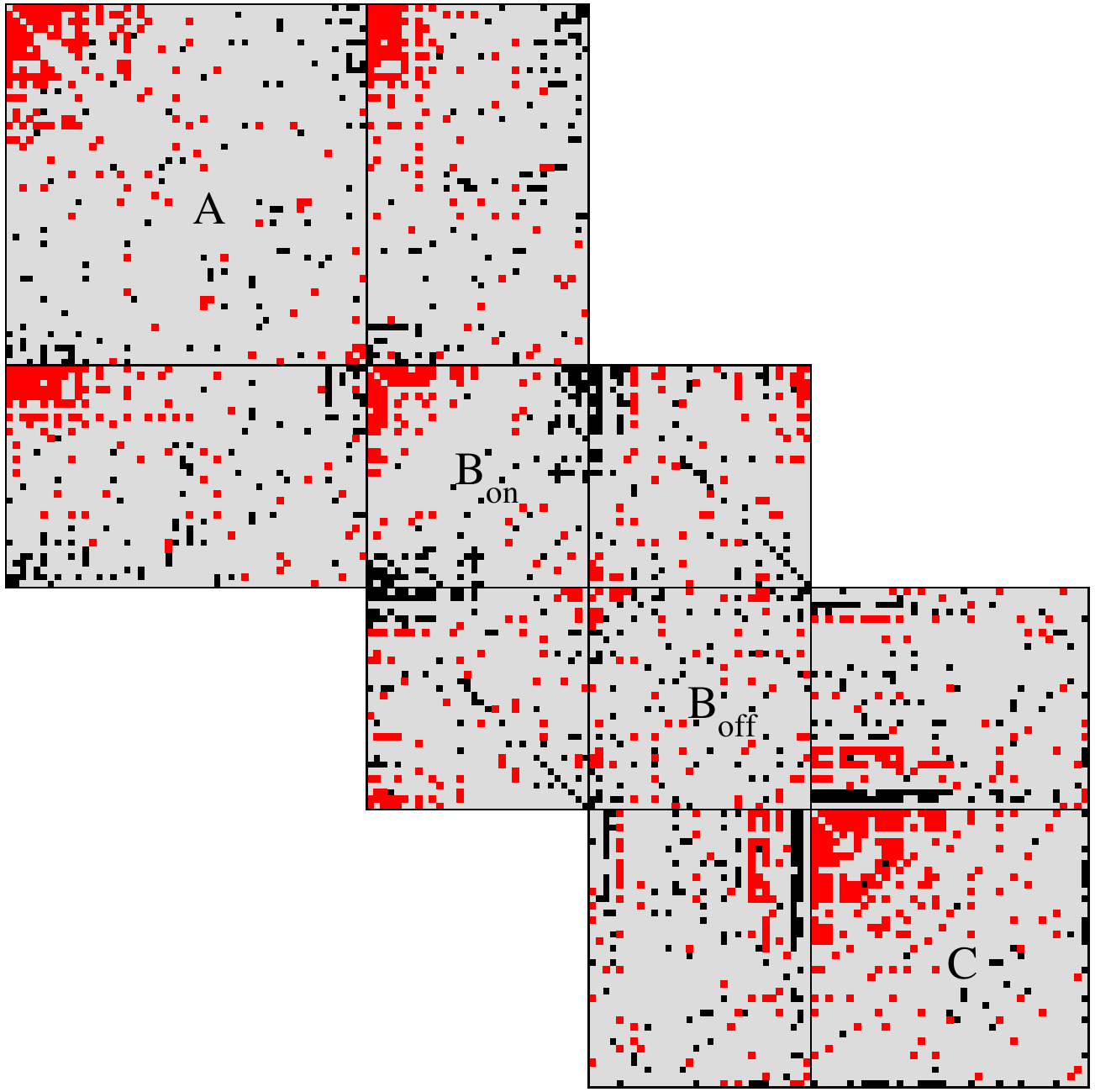}
\caption{\small{Color code representation of cross-correlation matrices for daily inventory variations of market members of set A, of market members of set B when considering the on-book market segment inventory variation ($B_{on}$),  of market members of set B when considering the off-book market segment inventory variation ($B_{off}$) and of set C. In the figure we also show the correlation coefficients between inventory variations $B_{on}$ and $B_{off}$, of sets A and $B_{on}$ and between sets C and $B_{off}$. The correlation matrix of market members of set A is ordered according to the value of correlation of on-book inventory variations of market members with the daily return of LLOY, ranking correlation from negative to positive values. The correlation matrices of market members of sets  $B_{on}$ and $B_{off}$ are ordered in a similar way. In the case of market members of set C the correlation used to rank the market members is the cross-correlation of off-book inventory variation with the daily return of LLOY from negative to positive values. Red (black) squares indicate positive (negative) values of cross-correlation which are statistically significant at a $2\sigma=0.1257$. The grey square spots indicates values of the correlation coefficient which are not statistically significant at the chosen threshold.}
}
\label{LLOYALL}
\end{figure}


In conclusion we find that in both market segments of LSE price returns is an important factor allowing to classify market members according to the response to return dynamics of their inventory variation. Some members, termed {\it trending} in Ref. \cite{Lillo2008}, are significantly positively correlated with price return, meaning that they buy when the price increases and sell when the price decreases. Other members, termed {\it reversing} in Ref. \cite{Lillo2008}, behave systematically in the opposite way, buying when the price goes down and selling when the price goes up. Inside each of these two groups, members are positively correlated while members of one group are typically anticorrelated with those of the other group.

A way of visualizing this interpretation is through the color code representation of the whole correlation matrix. The structure is quite complex because we have three sets of members (A, B, and C), two inventory variation profiles for group B (we term the on-book inventory variation of members in set B with $B_{on}$ and the off-book inventory variation of members in set B with $B_{off}$).
In Figure \ref{LLOYALL} we show a composition of cross-correlation matrices for the stock LLOY. We consider the cross-correlation matrix for the on-book inventory variations of set A, for the off-book inventory variations of set C, for the on-book inventory variations of set B ($B_{on}$), for the off-book inventory variations of set B ($B_{off}$) and the intersections between sets acting in the same market segment. We also show the correlation between the on-book and off-book inventory variation of market members active in both market segments (set B). The color code is the following: Red squares indicate positive cross-correlations ($\rho$) at a  $\rho > 2\sigma=2/\sqrt{253}=0.1257$ statistical significance threshold, black squares indicate negative cross-correlations at a  $\rho<-2\sigma=-0.1257$, while grey squares are correlation coefficients whose values are not statistically significant at the chosen threshold. Each matrix (with one exception, see below) is sorted according the cross-correlation between the market members inventory variations (electronic or off-book depending on the set considered) and the price return from the smallest negative value to the biggest positive value. The set $B_{off}$ is sorted as the set  $B_{on}$ because we wish to put emphasis on the correlation between the on-book and off-book inventory variation of the same members active in both market segments\footnote{When we sort according the cross-correlation values between the price return and the off-book inventory variation we obtain a similar structure.}.

From the panels of figure  \ref{LLOYALL} we make several observations. First, in the upper left corner of all the matrices around the main diagonal a large red area is typically observed. This corresponds to the group of reversing members, which are anticorrelated with returns and strongly correlated among them. As already observed at BME \cite{Lillo2008}, the group of reversing is much more strongly cross correlated than the group of trending and this is true for almost all the considered subsets with the possible exception of $B_{off}$.
The off diagonal matrix between set A and set $B_{on}$ displays a positive correlation between reversing of the two groups, while the off diagonal matrix between set C and set $B_{off}$ does not display such an area. In other words, on-book the reversing group forms a very correlated set indicating that  reversing of group A and $B_{on}$  trade in the same periods of time. In the off-book market  set $B_{off}$ does not show a strong reversing community but there is some correlation with the group of reversing of set C. This is clearer when the matrix $B_{off}$ is sorted according to the cross-correlation values between the price return and the off-book inventory variation (data not shown).

Market members who are very active in both market segments (set B) tend to take opposite position in the two segments. This is evident from the black dots (corresponding to significant negative correlations) in the diagonal of the correlation coefficient matrix observed between  $B_{on}$ and  $B_{off}$. To be more quantitative the average correlation between the on-book and off-book inventory variation of market members in set B is $-0.38\pm 0.31$ for LLOY and $-0.37\pm0.30$ for the whole sample of stocks. These members therefore transfer shares from one segment to the other. In other words they can buy (sell) from a non-member in the off-book and then sell (buy) in the on-book or can alter their inventory variation by buying (selling) in the on-book and therefore sell (buy) a typically larger amount in the off-book to a large client.

\begin{figure}[Ht]
\includegraphics[scale=0.5]{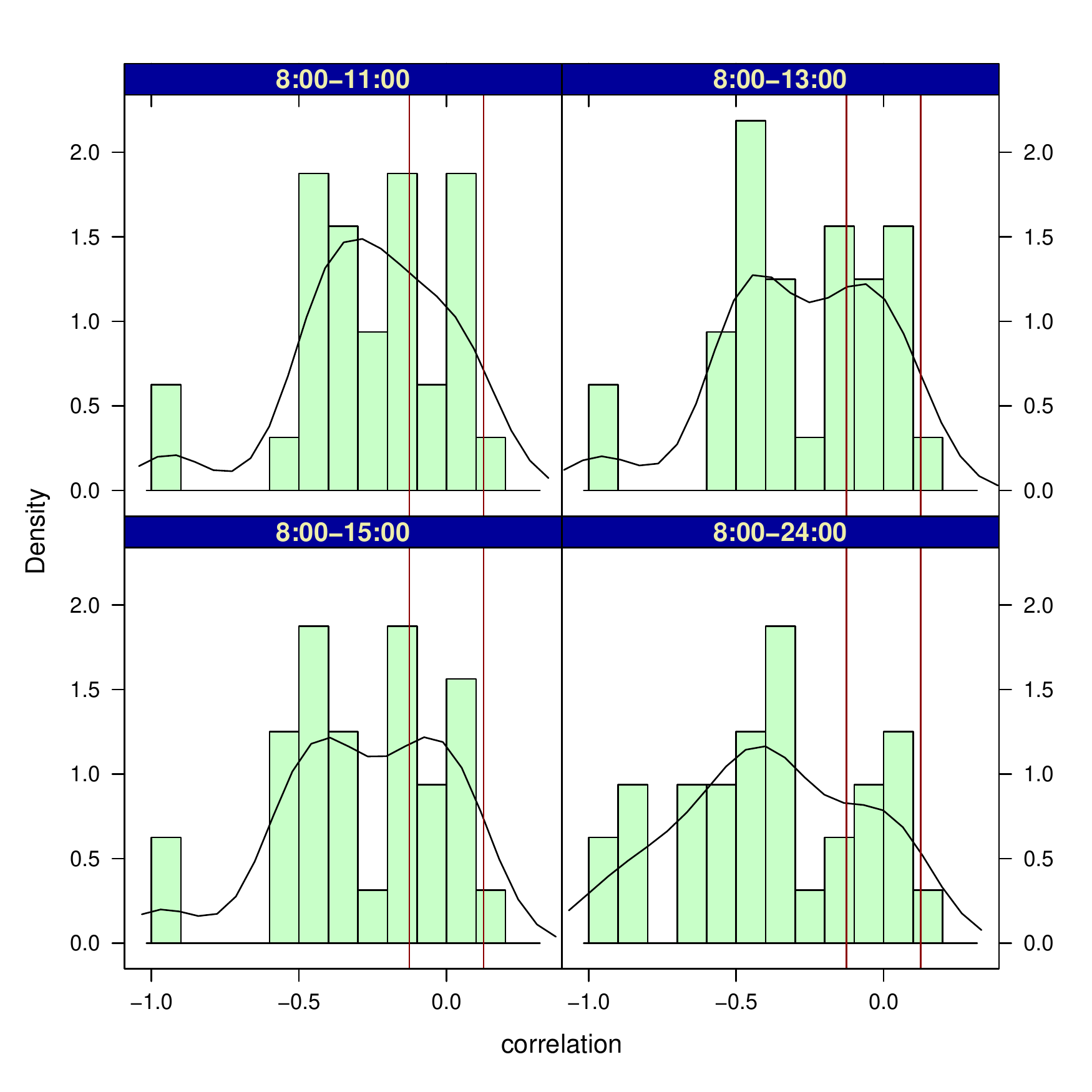}
\caption{\small{Histogram of the correlation coefficients  between on-book and off-book inventory variation of the same market member acting in both market segments. Different panels refer to time series of inventory variation computed within a different time window of the trading day. We show histograms of the correlation coefficient of inventory variation time series computed in a time window ranging from the opening of the market (8:00 in GMT time) until 11:00 (top left panel of the figure),   13:00 (top right),  15:00 (bottom left) and 24:00 (bottom right). In the four panels we also show as a continuous line a curve smoothing and interpolating the density function of correlation coefficients. The two vertical lines identify to the $2\sigma=0.1257$ confidence interval.}}
\label{TimeRho}
\end{figure}

The time profile of this kind of activities, i.e. acting as a brokerage house or acting a dealer of a client can be monitored by investigating the correlation between on-book and off-book inventory variation of the market members active in both market segments (set B). In Figure \ref{TimeRho} we show the histogram of the correlation coefficients between on- and off-book inventory variation of the same market member acting in both market segments  computed for different periods of the trading day. Specifically, we consider inventory variation time series for the transactions occurring from the opening of the market (8:00 in GMT time) until 11:00 (top left panel of the figure),   13:00 (top right),  15:00 (bottom left) and 24:00 (bottom right). In the four panels we also show as a continuous line a curve smoothing and interpolating the density function. The profile of the histogram and of the smoothing curve show that the on-book off-book inventory variation correlation assumes values which are more and more negative during the elapsing of time inside each trading day supporting the interpretation that market members of set B are specialized in the interaction with the two venues of the market and act as the agents processing trading activity that need to be transferred from one venue to the other.

\section{Trading activity of non-members: the correlation of order flow}
Non-members are allowed to trade only in the off-book market. With our database we are unable to discriminate individually non-members because many investors are classified by the same code. For this reason we decide not to investigate the trading activity of individual codes associated to non-members. On the other hand, we believe that the aggregate profile of all non-member participants can be informative under several aspects. Here we consider the statistical properties of the flow of orders\footnote{Note that while in the on-book market one order can generate more than one trade, in the off-book market we observe trades and we infer the order that generated it.} generated by non-members and traded with members (we have seen in Table I they represent the vast majority of trades generated by non-members). Part of this flow of trades is routed by members in the on-book and part is managed off-book by the market members, either by being a counterpart or by searching for another counterpart off-book.

\begin{figure}[Ht]
\vskip 1cm
\centering
\includegraphics[scale=0.6]{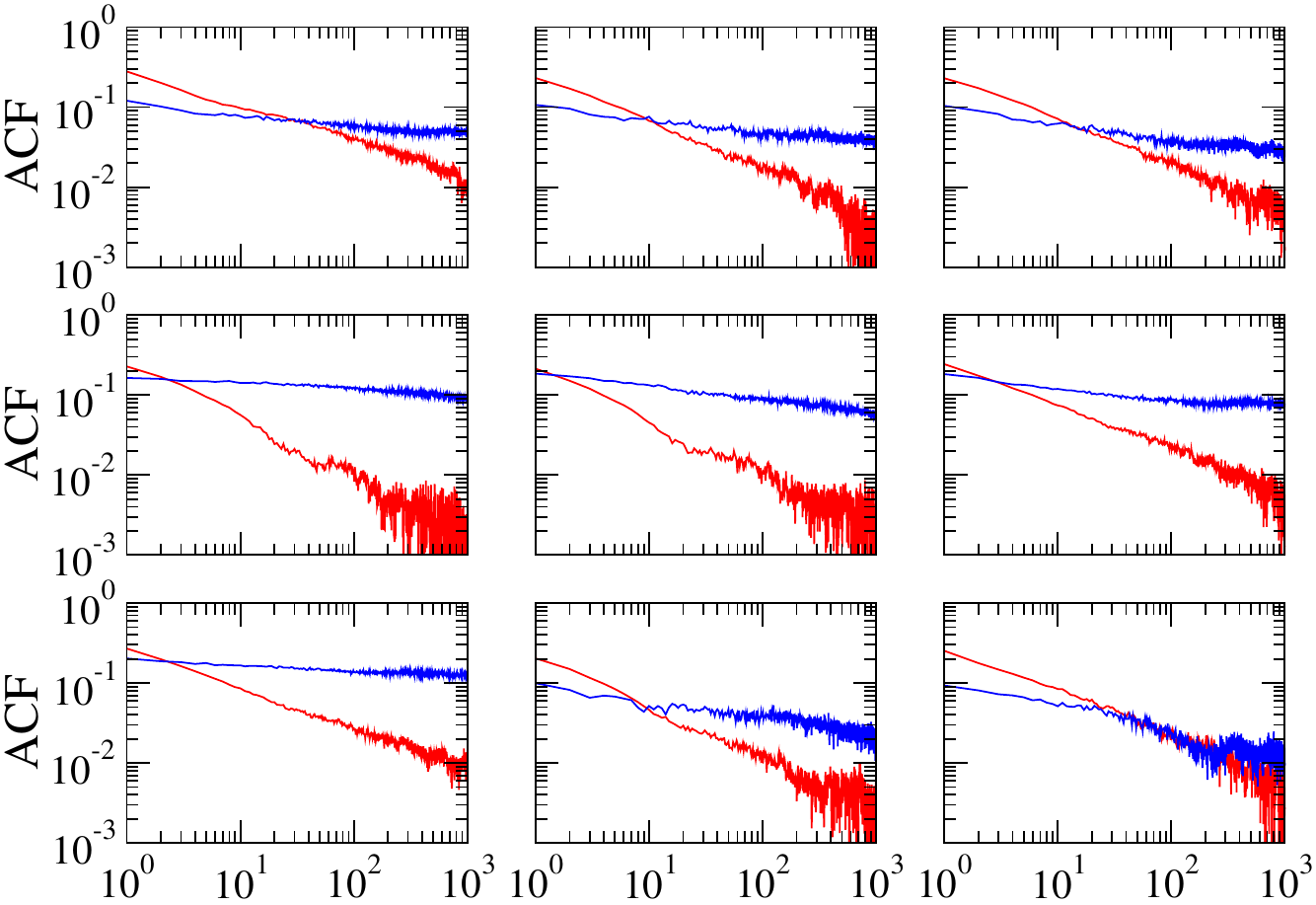}
\caption{\small{Each panel shows the autocorrelation of signs of market order flow in the on-book market (in red) and of the orders of non-members in the off-book market (in blue) for the 9 stock (from the top left to the bottom right the stocks are: VOD, GSK, BP, LLOY, RBS, HSBA, SHEL, AZN, BT-A).}}
\label{acf}
\end{figure}

It has been recently showed that the flow of the type (buy or sell) of market orders, i.e. orders triggering a trade in the limit order book, is highly correlated in time \cite{Lillo2004,Bouchaud2004}. Specifically, the autocorrelation of the time series of the sign $\epsilon_t$ of market orders decays as a power law function
\begin{equation}
\rho(\tau)=E[\epsilon_t \epsilon_{t+\tau}]\sim\frac{C}{\tau^\alpha}~~~~~~~~0<\alpha<1
\label{longmem}
\end{equation}
where $C$ is a constant.
Time series with this properties are termed long-memory processes \cite{Beran94}.
Long memory processes can be characterized by the exponent $\alpha$ describing the asymptotic behavior of the autocorrelation function or equivalently in terms of the Hurst exponent $H$ that, for long memory processes, is $H=1-\alpha/2$. Long memory processes are an important class of stochastic process that have found application in many different fields. The autocorrelation function of a long memory process is not integrable in $\tau$ between $0$ and $+\infty$ and, as a consequence, the process does not have a typical time scale. The fact that market order flow is long memory can be explained by herding among market participants \cite{LeBaron} or by splitting of large orders \cite{lmf2005,palit2011} and has important consequences for price formation and market efficiency (for a review see \cite{bfl}).

Here we compare the autocorrelation of signs of market orders placed in the on-book market by members to the autocorrelation of the sign of the trades in the off-book market placed by non-members. More precisely, for each off-book transaction between a member and a non-member, we assign the sign $\epsilon_t=+1$ ($-1$) if the non-member was the buyer (seller). This time series and the corresponding autocorrelation gives a measure of the time clustering of the order flow of non-members. We consider signs rather signed volume because absolute volume can be very heterogeneous across non-members (for example, retails vs institutions), while we are interested in the time clustering of trading intentions.

Figure \ref{acf} shows the autocorrelation of signs both on-book and off-book  for the nine investigated stocks. We observe that on-book market order sign time series is well described by a long memory process according to Eq. (\ref{longmem}).  Quite surprisingly, the time series of signs of off-book trades by non-members is also very correlated. The autocorrelation function starts at a value smaller than the one for on-book trades, but the off-book autocorrelation crosses the on-book autocorrelation and decays very slowly. This suggests that on short time scale off-book trades are less correlated than on-book trades, while on longer time scales the opposite is true.

\begin{figure}[Ht]
\vskip 1cm
\centering
\includegraphics[scale=0.6]{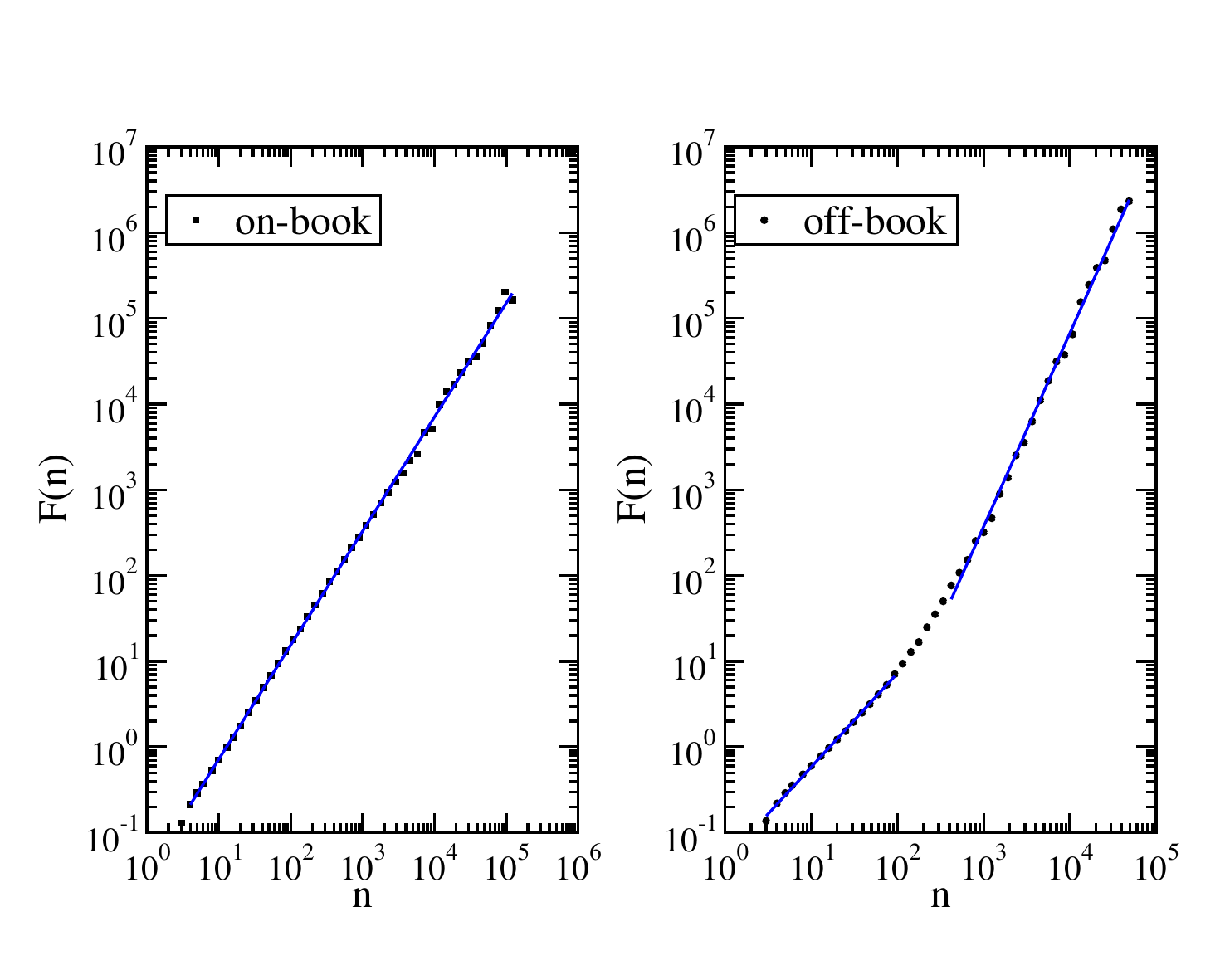}
\caption{Detrended Fluctuation Analysis of the time series of market order sign in the on-book market (left) and of the time series of the sign of transactions performed by non-member in the off-book market (right) for the stock LLOY. The lines are best fit of power law functions. See the text for the exponents.}
\label{DFA}
\end{figure}

To quantify this effect we compute the Detrended Fluctuation Analysis \cite{dfa} of the two time series for each stock. This method was introduced more than fifteen years ago to originally investigate molecular biology symbolic sequences. Since its introduction it has been applied to a large variety of systems, including physical, biological, physiological, economic, and technological data. The idea is to consider the integrated process and detrend it locally. The scaling of the fluctuations of the residuals as a function of the box size in which the regression is performed gives the estimate of the Hurst exponent.

Figure \ref{DFA} shows the  fluctuation of the residuals as a function of the box size. For a pure long memory one should observe a power law dependence with an exponent equal to two times the Hurst exponent. We performed the DFA for the on-book and the off-book transaction sign time series. It is clear even by eye the difference between the two DFA plots. For on-book data a single power law line is able to capture the whole range of time scales. The Hurst exponent obtained from a least square regression of on-book data is $H=0.67$. On the other hand the DFA plot of off-book data shows clearly at least two different  regimes. A power law fit in the region corresponding to box sizes smaller than 100 transactions gives a local Hurst exponent of $H=0.55$, very close to a short memory process. A power law fit in the region corresponding to box sizes larger than 500 transactions gives an Hurst exponent of $H=1.13$ (see Fig. \ref{DFA}).

\section{Price impact in the two markets}

We now turn to the investigation of the impact of transactions on price. Price impact is the expected price change conditioned on initiating a trade of a given size and a given sign.  Understanding market impact is important for several reasons, including the control of execution costs and the understanding of how information is transferred to price (for a recent review see \cite{bfl}). Price impact depends in general on many variables, including volume and time. Here we want to quantify the temporal and volume structure of price impact of individual transactions by performing a comparative analysis between on-book and off-book trades. There is a clear difference between the impact in the two market segments. In the on-book segment, market orders have a mechanical component of the impact due to the fact that, if the market order size is larger than the volume at the opposite best, the midprice will automatically move. On the contrary off-book transactions do not have a mechanical impact on on-book prices. The impact of an off-book trade is informational or it can be due to an indirect effect. For example if a non-member trades off-book with a member, it is possible that the same member trades the generated inventory imbalance in the on-book market, creating an effect on price.

\begin{figure}[Ht]
\vskip 1cm
\centering
\includegraphics[scale=0.6]{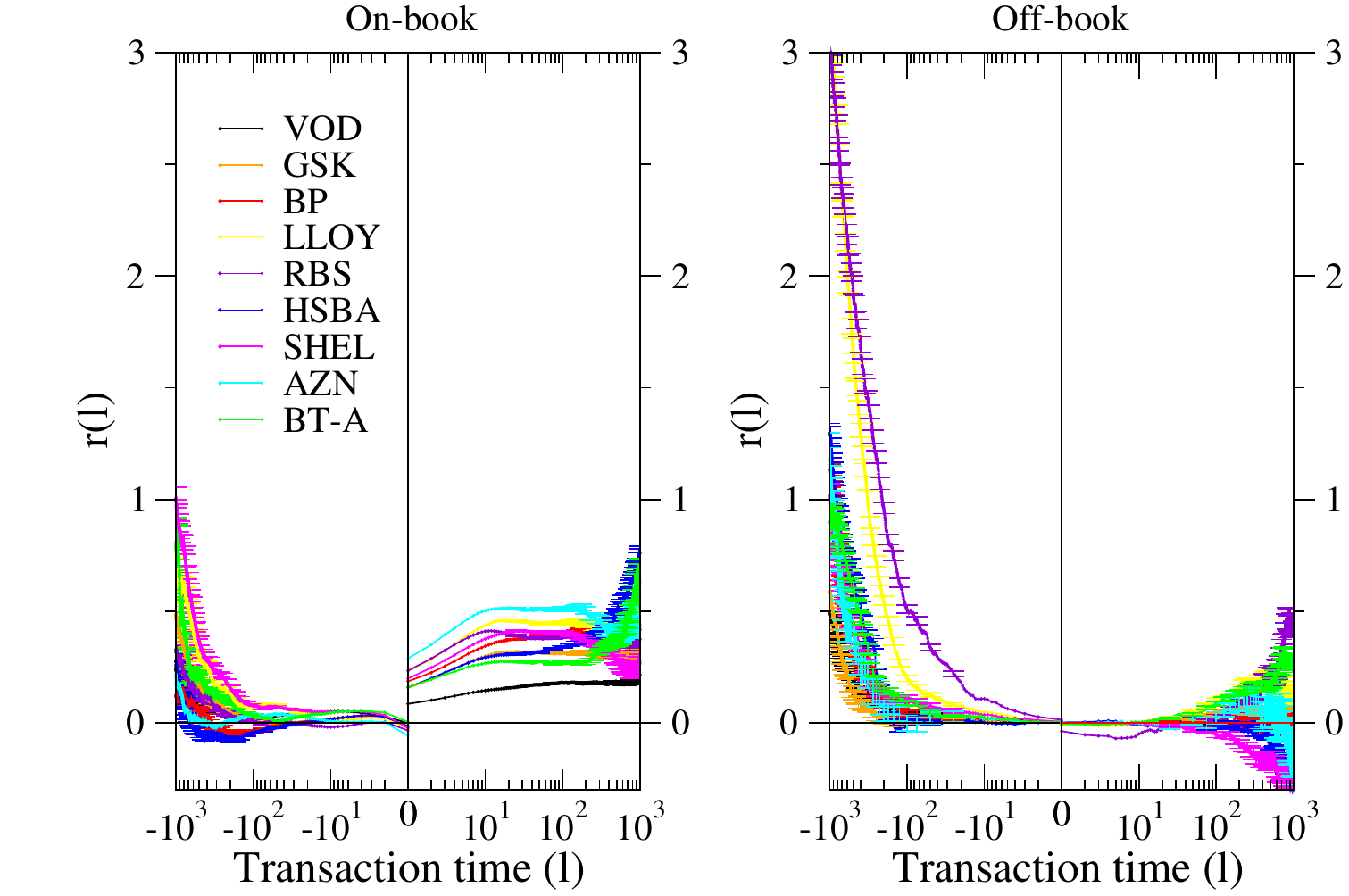}
\caption{\small{Price response function (see Eq. \ref{Bou}) for on-book trades (left) and for off-book trades (right) for all the nine investigated stocks in the 2004. The price response function is in spread units and time is in on-book transaction time.}}
\label{ImpactOnOff}
\end{figure}

In order to study the temporal component of price impact we compute the response function ${\cal R}(l)$ introduced in \cite{Bouchaud2004} and defined for on-book transactions as
\begin{equation}
{\cal R}(l)=E[(p_{n+l}-p_{n})~\epsilon_n]
\label{Bou}
\end{equation}
Here $p_n$ is the (log) midprice  immediately before the trade at on-book transaction time $n$ occurs and $p_{n+l}$ is the (log) midprice immediately before a transaction at a lagged on-book transaction time $n+l$. Finally, $\epsilon_n$ is the sign of the market order initiating the trade at time $n$.
For off-book trades we consider as before only transactions between a member and a non-member and $\epsilon_n$ is the transaction sign of the non-member.
Differently from previous analyses we consider both positive and negative values of $l$, ranging in the interval $l \in [-1000,1000]$. The response function for negative lags $l$ tells us how the price was moving in the past conditioned to the fact that we are now observing a buyer or a seller initiated transaction. In order to compare different stocks we compute the response function in unit of spread, obtained as an unconditional average across the whole 2004. Spread is also a natural unit for measuring impact of individual transactions \cite{wyarth}. We compute error bars as standard errors.
To avoid overnight effects we consider only events at time $n$ and $n+l$ that occur in the same trading day.  We also systematically discard the first and last fifteen minutes of trading in each day to remove effects due to the opening and closing of the market.

Figure \ref{ImpactOnOff} shows the response function in on- (left) and off- (right) book for the nine stocks. For on-book transactions the response function increases with time and then flattens out as already observed in \cite{Bouchaud2004}. For negative lags we observe a smaller, but positive response. If statistically significant, this implies that by observing a buyer initiated on-book transaction now, one can conclude that on average price was declining in the recent past. This effect is much stronger for off-book transactions. In fact, as shown in the right panel of  Figure \ref{ImpactOnOff}, if a non-member buys off-book now, on average price was strongly declining and vice versa for sell trades. This suggests a strong contrarian behavior of non-members. Note also that the impact for positive lags is negligible. This is not due to the fact that off-book transactions do not have a mechanical component of the impact. In fact even subtracting the $l=1$ component from the response function, the on-book transaction response grows in the future, while the off-book transaction response is essentially consistent with zero.

\begin{figure}[Ht]
\vskip 1cm
\centering
\includegraphics[scale=0.6]{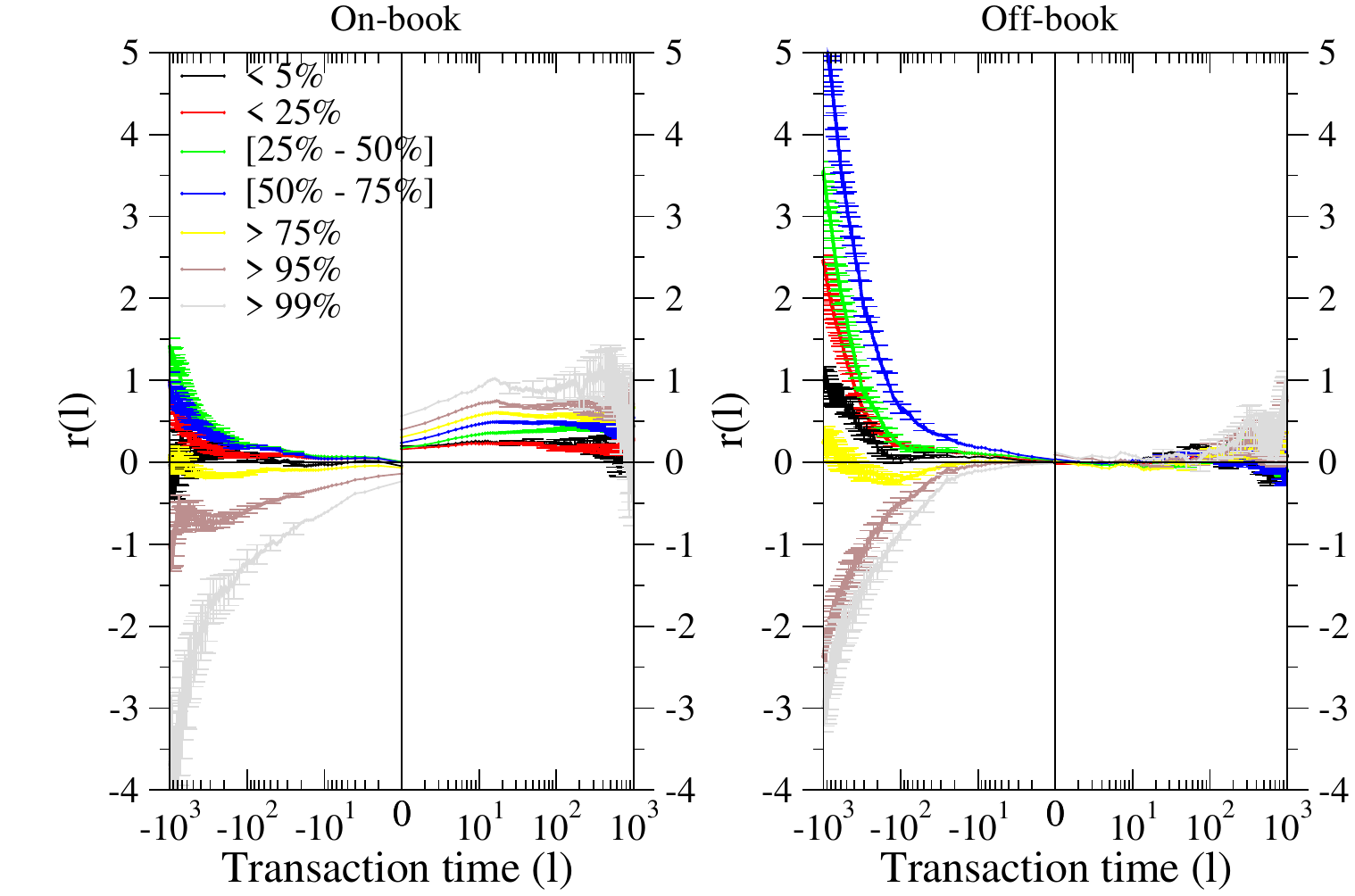}
\caption{\small{Price response function  (see Eq. \ref{Bou}) for the on-book trades (left) and for off-book trades (right) for the stock LLOY in 2004. Different colors refer to different percentiles of transaction volume. In particular: in black there are transactions whose volume is lower that the 0.05 percentile, in red the volume is lower than the 0.25 percentile, in green the volume is between 0.25 and 0.50 percentile, in blue the volume is between the 0.50 and 0.75 percentile, in yellow the volume is greater than the 0.75 percentile, in brown the volume is greater than the 0.95 percentile, and finally in grey the volume is greater than the 0.99 percentile. The price response function is in spread units and time is in on-book transaction time.}}
\label{ImpactOnOffVol}
\end{figure}

This behavior is obtained by pooling together all the transactions independently on the volume. Authors of Ref. \cite{Bouchaud2004} mentioned that when conditioning on the volume of the on-book transactions, the response function roughly factorizes in a lag dependent part and a volume dependent part, which is essentially the immediate impact of individual transactions as a function of volume. In order to compare on- and off-book transactions we compute the quantiles of transaction volume of the pooled sample of on- and off-book trades. We then compute the response function in the two market segments conditioned to fact that the volume of the transaction at time $n$ belongs to a given volume quantile.

We find that the behavior of the price response function is strongly dependent on the volume of considered transactions, both for the on-book and the off-book trades.  Figure \ref{ImpactOnOffVol} shows the response function conditioned to volume for LLOY. For on-book trades, we observe a clear monotonic dependence of response function for positive lags as a function of volume, as mentioned in \cite{Bouchaud2004}. For negative lags the on-book impact is slightly positive for most quantiles, except for the top quantiles (above 95\%). For these large transactions the response is large and negative, meaning that a large buyer initiated trade is typically preceded by a price increase. A possible explanation is related to order splitting. Large transactions are likely related to large orders that are split and traded incrementally over extended periods of time. This process pushes the price in the direction of the trade (especially if the order is executed through market orders) and this could explain the negativity of price response of large volumes for negative times. The dichotomy between large and small trades is even more evident for off-book trades (right panel of Fig. \ref{ImpactOnOffVol}). Large trades have a negative response for negative lags (as for on-book trades), while smaller trades have a large positive response that again could be interpreted as a contrarian behavior. There is strong empirical evidence that individual investors (retail) have preferentially a contrarian trading strategy
\cite{Nofsinger1999,Grinblatt2000,Barber2008,Grinblatt2009}.

\section{Conclusions}
We have presented empirical results on the trading profile of active market participants of the London Stock Exchange. Specifically, we investigated market members trading in the electronic on-book market and/or in the dealership off-book market. Market members interact among them in the on-book market and with other market members and/or with non-members in the off-book market. A primarily goal of the present study was to investigate similarities, differences and  the interplay between the trading decision taken by the market members in the on-book and in the off-book segments of the market. We find several statistical regularities
showing that the presence of different market venues play a role in shaping the trading activity of market members of a stock exchange. A first result shows that the exchanged value probability density is strongly affected by the market venue where the transactions occur. In fact, the probability density function of exchanged value in a off book transaction has a much fatter tail than the the probability density function for on-book trades. The daily trading profile of market members which are active in both segments of the market shows that  a strong anticorrelation is present between the inventory variation of a member built up from the on-book market transactions and inventory variation built up from the off-book market transactions with non-members. This result indicates that this set of market members act in the market providing a bridge between the trading activity performed in the two market segments.

Another result obtained concerns the memory of the sequence (categorized as +1 for a buy and -1 for a sell transaction) of transactions observed in the two segments of the market. The autocorrelation of the on-book and off-book time series is quite different. In the case of the on-book market segment we confirm the known power law behavior originally proposed in \cite{Lillo2004,Bouchaud2004} whereas in the off-book market segment the decay of the autocorrelation function presents two distinct regimes: (i) a fast decay regime and
(ii) we a very slow decay for long times. This result suggests that the role of retail investors and institutional investors might be different with respect to the nature of the order flow they generate.

Finally we analyze the price impact function observed in the order book both for positive and negative time lags. The investigation is done by considering the effect of transactions of both market segments on the evolution of the the mid-price of the order book. This is done for the on-book and the off-book transactions  both unconditionally, averaging over all on-book or off-book trades, and conditioning the average market impact over some percentiles of transaction value. We find that the unconditional curves are very different for the on-book electronic transactions and for the off-book trades. For the unconditional case, market impact of the on-book transactions is quite evident whereas off-book transactions seems to play no role. We find also a little dependence on the value of the transaction for the on-book transactions while, on the contrary, really a big difference in the values and in the shape of the lagged market impact (both for positive and negative lags) for the off-book transactions when we condition on the the percentile of the transaction values.

\section*{Acknowledgments} Authors acknowledge financial support from the MIUR PRIN project 2007TKLTSR  ``Indagine di fatti stilizzati e delle strategie risultanti di agenti e istituzioni osservate in mercati finanziari reali ed artificiali".

\end{document}